\documentstyle[12pt]{article}
\begin{document}
\vskip 4 cm
\begin{center}
\Large{\bf An Alternative Framework of Geometry and Topology in 
Relativity}
\end{center}
\vskip 2 cm
\begin{center}
{\bf Syed Afsar Abbas} \\
Centre for Theoretical Physics\\
JMI, New Delhi - 110025, India\\
(e-mail : afsar.ctp@jmi.ac.in)
\end{center}
\vskip 15 mm  
\begin{centerline}
{\bf Abstract }
\end{centerline}
\vskip 3 mm

We study geometry and topology as complementary and dual aspects of
the mathematical space. The same is used to get a better understanding of 
the Cosmological Constant. Having failed so far to include gravity in a 
proper unified framework with the other three fundamental gauge forces, 
we are now faced with an additional unwanted fifth force of repulsion, also 
envisaged as the Dark Energy problem. How does one understand this 3+1+1 
fundamental force dilemma? We introduce here a novel idea of the Fundamental 
Forces. This will give us an additional and an all-encompassing way of 
classifying these five fundamental forces in a consistent manner
and thereby strengthen the geometry-topology complementarity concept.
This also helps us to understand as to what one actually means when 
substituting ${g_{\mu \nu}} = {\eta_{\mu \nu}} + {h_{\mu \nu}}$
with the last term being "small" in General Relativity. This provides an 
understanding of the generic relationship and the complementarity of the  
geometric and the topological structures which is then formalized as a 
basic theorem enabling us to understand the underlying connection 
between the physical reality and the pure mathematical structures. 

\newpage

To understand the structure of space-time, one starts with the simplest 
possible differential manifold as a collection of points. One imposes 
conditions of smoothness on it to define the most primitive topological
structure on it. These topological transformations are quite independent 
of the concept of length. Thereafter one introduces a new property on 
the manifold, that of an affine connection and also introduce a new
structure which brings in the notion of length through a metric.
In general, the affine connection concept and the metric are quite 
independent of each other. However, as it turns out that in physics what 
appears to have physical relevance is an affine connection defined in 
terms of the metric - the so called metric connection. 
One should note that in this metric structure, the topological structure 
is anyway present simultaneously in the background - meaning that 
both the geometric and the topological structures are present 
intrinsically.

We have also learned that metric or length has all that can be known 
about the structure of the space-time. If we know the metric we have
learned as much as possible about space-time. 
But one should not forget the 
significance of "no metric". It should have physical significance too.
This corresponds to the topological structures. 
Invariants of the topology
provide further basic information about the spacetime structure.
So geometry ( properties dependent upon the existence of a metric ) and 
topology ( properties independent of length ) seem to be playing a 
complementary role here. This is further consolidated by the fact that
when we deform a surface, properties of the surface which do not change 
with deformation are called topological while those which do change as a 
result of deformation are geometric. 
Not only does it appear that the topological 
structures and geometric structures are complementary aspects of the 
mathematical space; in as much as a metric may be present or not present - 
the complementarity is exclusive, 
and in as much as these are the only two possibilities
which a physically relevant structure can have - the complementarity
is exhaustive. Viewed in this manner,
this appears like a fundamental structural duality of space-time.
One should note that this duality is of generic nature, not dependent 
upon the details of either geometry or topology involved.

The question that arises is that as what appears as a logically 
and mathematically consistent connection
between geometry and topology above, does the physical reality as 
emphasized in General Theory (GT), Special Theory (ST) 
and Newtonian Mechanics (NM) reflect this structural duality?
We study this below to show that indeed there are strong physical support 
for this concept. This enables us to suggest a new 
Geometry-Topology-Complementarity Theorem to formalize this structure and 
which in return pays back rich dividends.

Note that we may define geodesics as per Wald [1, p 41] as a curve whose 
tangent vector is parallel propagated along itself, which means a curve 
whose tangent $T^a$ satisfies the equation

\begin{equation}
T^a {\nabla_a} T^b = 0
\end{equation}

where $\nabla_a$ is the covariant derivative operator. This definition may 
be called "geometric" in nature as we are demanding maintenance of the
same length. However we may impose a weaker condition on parallel 
transport that the tangent vector to the curve point in the same direction 
as itself when parallel transported without demanding any maintenance of the 
same length.
In that case the above condition becomes:

\begin{equation}
T^a {\nabla_a} T^b = \alpha {T^b}
\end{equation}

with $\alpha$ being an arbitrary function on the curve.
We call this geodesics as being topological in nature as the
concept of length does not arise here. However it turns out that
the second equation can be reparametrized ( affine parametrization ) such 
that the first equation arises. And there is no loss of generality in 
doing this.
In terms of the complementarity principle that we have enunciated above, 
this means that both the geometric and the topological characters are 
hidden within the structure of GR without favouring any one particular 
aspect - geometry or topology. 
The geodesics are neutral as to this - but they are 
both there intrinsically.
However when GR will make predictions for physically measurable 
quantities, one feels that this will provide it to split 
with a geometric  and a complementary ( that is additive ) 
topological character. Let us see how this may be justified.

Einstein Equation says

\begin{equation}
G_{\mu \nu} = 8 \pi G { T _{\mu \nu}}
\end{equation}

Due to the Equivalence Principle, on the left hand side the force
of gravity has disappeared entirely and has been replaced by pure
geometry. The Machian view is reflected in the energy momentum tensor 
on the right determining the geometric structure on the left. 

This equation has been very successfully employed to understand the 
structure of space-time and it has stood well in giving good
understanding to cosmology.
A very successful equation indeed, until recently when one learned that
actually there is a new repulsive force within the framework of the
expanding universe. How does one understand that?

Let us look at the Einstein's Equation again. 
Harvey and Schucking correcting for Einstein's error [2]
in mis-understanding the role of the cosmological term $\lambda$,
have derived the most general equation of motion to be

\begin{equation}
G_{\mu \nu} + \lambda {g_{\mu \nu}} = 8 \pi G {T _{\mu \nu}} 
\end{equation}
 
They showed that the Cosmological Constant $\lambda$ above provides 
a new repulsive force proportional to mass m, repelling every 
particle of mass m with a force

\begin{equation}
F = (m \lambda) {{{c^2} x} \over 3}
\end{equation}

Hence so to say there is "matter without motion" [2]
where Cosmological Constant provides the repulsion.
As per our complementarity principle between geometry and topology,
and as geometry has already replaced the force of gravity above, 
could it be that this additional repulsive force be treated as of topological 
nature. The fact that it is force without motion, implies that there
is no length concept involved. We have seen above that mathematical 
structures which are independent of length are of topological nature. So 
indeed this new force may justifiably be treated as being of topological 
nature. So we notice that geometry-topology complementarity
theorem seems to be holding good here. 

The same concept here gets support from a different quarter - that of 
consistent understanding of as to how many fundamental forces actually are 
there; four or five or more?

So far all our understanding of nature has been successfully
described within the Standard Model (SM) of particle physics.
Whatever was not accessible to it, has been explained in terms of various
theoretical extensions of the SM.
All this was done in terms of an understanding that there are four
fundamental forces. Three of these are gauge forces and the fourth one, 
that of gravity, it is believed, shall "soon" be incorporated in a
unified whole as some kind of quantized gauge theory. 
This "soon" has been dogging us for several decades.
The problem becomes more confusing in that there always remains a 
clear possibility that gravity, at a fundamental level, may 
be a different kind of force altogether and may not be quantized
at all, and in which case its unification with the other three forces 
will have to be seen differently.
The fact that one has not been able to achieve this so called unification
of the four forces so far, we
are thus justified in breaking this so called four force problem as 
actually being of the nature of a 3+1 force problem. 

Given the above situation,
no one expected and no one wanted, yet another new 
fundamental "force" to spring up.
But there it is - the new force of repulsion of galaxies given by eqn. 5,
call it RF (Repulsive Force)! 

One question that arises immediately is, as 
to the nature of this RF. Is it a simply a gauge force like the other 
three and then the force problem is of the 4+1 kind; or 
is it fundamentally of the gravity kind and in which case the force 
problem is that of 3+2 kind; or is it different from all these 
and in which case it is 3+1+1 kind? 

The discovery of the RF is akin to the discovery of the muon, 
when people were 
quite happy and contended with only the electron and when I. I. Rabi 
in puzzlement asked, "who ordered it?" 
We too can paraphrase Rabi by asking, "Who ordered this fifth force?" 
The discovery of muon forced scientists to extend their theoretical 
framework significantly.
No patch-up work, but a genuine attempt to include this new force
in a fundamental and consistent framework of our understanding of 
nature.  

It may be remarked that the concept of a so called  
fifth force has been there for quite 
sometime. Extensions of Einstein's GTR, like for example Brans-Dicke 
theory, necessarily have an extra fifth force, in which case the RF may 
belong to the 3+2 or 3+1+1 classification. 
Higher dimentional
Kaluza-Klein kind of theories, supersymmetric theories, superstring 
theories etc also predict the fifth fundamental force of the Yukawa kind 
and in which case it will very likely belong to the 4+1 kind. It is not 
clear that the new RF is this putative fifth force [3,4].
In fact this theoretical fifth force is incompatible with overall
cosmological framework [3,4].
Just because the word "fifth" force has been usurped by the other
models, does not mean that the actual empirical fifth RF is of their kind.
So minimal conclusion would be that with the new RF, the force problem is 
per se of the 3+1+1 kind. 

Here we wish to understand the "force" nature of the new problem.
To do so we introduce a 
new concept of the "Universal Force". It was first proposed 
by Hans Reichenbach [5]. 

Reichenbach defines two kind of forces - Differential Forces and 
Universal Forces. It may be pointed out that 
the term "force" here should not be taken strictly as
defined in physics but in a broad and general framework.

One calls a force Differential if it acts differently on different 
substances. It is called Universal if it is quantitatively the 
same for all the substances [5]. If we heat a rod of initial 
length $l_0$ from initial temperature $T_0$ to temperature T then 
its length is given as

\begin{equation}
l = l_0 [ 1 + \beta ( T - T_0 ) ]
\end{equation}

where $\beta$ the coefficient for thermal expansion is different 
for different materials. Hence this is a Differential Force.
Now the correction factor due to the influence of gravitation on 
the length of the rod is

\begin{equation}
l = l_0 [ 1 - C { m \over r } {cos{^2} \phi}]
\end{equation}

Here the rod is placed at a distance r from sun whose mass is m 
and $\phi$ is the angle of the rod with respect to the the line 
sun to rod. C is a universal constant ( in CGS unit 
C= 3.7 x ${10}^{-29}$ ). As this acts in the same manner for any 
material of mass m, gravity is a Universal Force
as per the above definition.

Reichenbach also gives a general definition of the Universal 
Forces [5,p 12] as: (1) affecting all the materials in the same 
manner and (2) there are no insulating walls against it. We saw 
above that gravity is such a force,

Indeed gravity is a Universal Force par excellence. It affects all 
matter in the same manner. The equality of the gravitational and 
inertial masses is what ensures this physically. If the 
gravitational and inertial masses were not found to be equal, then 
one would not have been able to visualize of the paths of freely 
falling mass points as geodesics in the four dimentional space-time.
In that case different geodesics would have resulted from 
different materials of mass points [5]. 

Therefore the universal effect of gravitation on different kinds 
of measuring instruments is to define a single geometry for all of 
them. Viewed this way, one may say that gravity is 
geometrized. "It is not theory of gravitation that becomes 
geometry, but it is geometry that becomes the experience of the 
gravitational field" [5, p 256]. Why does the planet follow the 
curved path? Not because it is acted upon by a force but 
because the curved space-time manifold leaves it with no other 
choice!

So as per Einstein's theory of relativity, one does not speak of 
a change produced by the gravitational field in the measuring 
instruments, but regard the measuring instruments as free from any 
deforming forces. Gravity being a Universal Force, in the 
Einstein's Theory of Relativity, it basically disappears and is 
replaced by geometry.

In fact Reichenbach [5, p 22] shows how one can give a consistent 
definition of a rigid rod - the same rigid rods which are needed 
in relativity to measure all lengths. "Rigid rods are solid bodies 
which are not affected by Differential Forces, or concerning which 
the influence of Differential Forces has been eliminated by 
corrections; Universal  Forces are disregarded. We do not neglect 
Universal Forces. We set them to zero by definition. Without such 
a rule a rigid body cannot be defined." In fact this rule also 
helps in defining a closed system as well. 

All this was formalized in terms of a theorem by 
Reichenbach [5, p 33]

\vskip 1 cm

{\bf THEOREM $\theta$} :

Given the geometry $G^0$ 
to which the measuring instruments 
conform, we can imagine a Universal Force F which affects 
the instruments in such a way that the actual geometry is an 
arbitrary geometry $G$, while the observed deviation from $G$
is due to universal deformation of the measuring instruments."

\begin{equation}
{G^0} + F = G 
\end{equation}

Hence only the combination ${G^0} + F$ is testable. 
As per Reichenbach's 
principle one prefers the theory wherein we put F=0.
If we accept Reichenbach principle of putting the 
Universal Force of gravity to zero, then the arbitrariness in the 
choice of the
measuring procedure is avoided and the question of the geometrical 
structure of the physical space has a unique answer determined by 
physical measurement. 

In the case of gravity, and in as much as Einstein's Theory of 
Relativity has been well tested experimentally, we treat the 
above concept as well placed empirically. But from this single 
success Reichenbach generalizes this as a fundamental principle 
for all cases where Universal forces may arise. 

As such Reichenbach goes ahead and tries to apply this principle 
of elimination of Universal Forces to another universal effect 
that he finds and which arises from considerations of 
topology ( as an additional consideration over and above that of 
geometry ) of space-time of the universe.

The Theorem $\theta$ is limited to talking about the geometry of
space-time only. It does not take account of specific 
topological issues 
that may arise. To take account of topology of the space-time 
we shall have to extend the said theorem appropriately.
This of course is essential as per our geometry-topology complementary 
concept.

What would one experience if space had different topological 
properties. To make the point home Reichenbach considers a 
torus-space [5, p 63]. This is quite detailed and extensive.
However for the purpose of simplifying the
and shortening the discussion here we shall
talk of a two dimensional being who lives on the 
surface of a sphere. His measurements tell him so. But in
spite of this he insists that he lives on a plane. 
He may actually do so as per our discussion above if he confines 
himself to metrical relations only.
With an appropriate Universal Force he can justify living
on a plane. But the surface of a sphere is topologically different 
from that of a plane. On a sphere if he starts at a point X and 
goes on a world tour he may come back to the same point X. But 
this is impossible on a plane. And hence 
to account for coming back to the "same point" 
he has to maintain that on the plane he 
actually has come back to a different point Y - which though is 
identical to X in all other respects. 
One option for him is to 
accept that he is actually living on a sphere. 
However if he still wants 
to maintain his position that he is living on a plane then he has 
to explain as to how point Y is 
physically identical to point X in spite of 
the fact that X and Y are different and distinct points of space. 
Indeed he can do so by visualizing a fictitious force as an 
effect of some kind of "pre-established harmony" [5, p 65] by 
proposing that everything that occurs at X also occurs at the 
point Y. As it would affect all matter in the same manner this 
corresponds to a Universal Force as per Reichenbach's 
definition.

This interdependence of corresponding points which is essential 
in this "pre-established" harmony cannot be interpreted as 
ordinary causality, as it does not require ordinary time to 
transmit it
and also does not spread continuously through intervening space. 
Hence there is no mysterious causal connection between the points 
X and point Y. Thus this necessarily entails 
proposing a "causal anomaly" [5, p 65].
In short connecting different topologies through a fictitious
Universal Effect of "pre-established harmony" necessarily calls 
for introduction of "causal anomalies". 
Call this new hypothetical 
Universal Force as A and the Theorem $\theta$ be extended to 
read

\begin{equation}
{G^0} + F  + A = {\bf G} 
\end{equation}

where on the right had side we have given a different 
capital ${\bf G}$ 
which reduces to $G$ of the original Theorem $\theta$
when A is set equal to zero.

Now as per Reichenbach's law of preferring that physical reality 
wherein all Universal Forces are put to zero, he 
advocates of putting A to zero. He pointed out that this has the 
advantage of retaining physical "causality " in our science.
 
However what it is possible that Reichenbach was 
wrong in putting all Universal Forces to zero. The justification has to be 
sought in actual physical reality. 
It was fine to put F to zero, which allowed us to define a truly
"rigid" rod and which led to a geometrical interpretation of 
gravity in a unique manner.
But in the case of this new topological Universal Force we 
really do not know enough 
and let us not be governed by any theoretical prejudice and 
let the Nature decide as to what is happening. So to say, let us 
look at modern cosmology to see if it is throwing up any new 
Universal Forces which may be identified with our 
"pre-established harmony" here.
And there indeed is this Universal RF here!
It is universal as it acts in the same manner on all
bodies of mass m. This new fifth fundamental force, which is a
puzzle for the SM and its putative extensions, is but a natural ally of 
gravity in being of universal character.

So as per this new classification, there are three well known gauge 
forces and two universal forces - that of gravity and the new one
of repulsion. 
However, this has an advantage that it points to a basic
similarity between the two - gravity and repulsive-force, 
which is not apparent in the canonical way of 
adding up the fifth force in an ad-hoc manner.
Hence as per the definition above, the forces should be 
classified as 3+2 kind. Clearly this is providing us with an understanding 
which may help us in the present puzzling scenario.

This thus supports our Geometry-Topology Complementarity concept.
And because of the exhaustiveness and the completeness of geometry and 
topology to define the surface completely,
it therefore predicts that there
would be no more fundamental forces other than the five known now.

One would like to ask as to in what other manner, incorporation of 
this new "causal anomaly", may help us in understanding Nature 
better? Will it provide new perspectives as 
answers to quantum mechanical puzzles of 
quantum jumps, non-locality etc. These are open questions  
to be tackled in future. 

As a further application of the Geometry-Topology Complementarity concept
let us study the Newtonian Mechanics (NM) limit of General Relativity 
(GR). One gets NM from GR under appropriate condition under
a perturbation approximation
with the assumption that 

\begin{equation}
{g_{\mu \nu}} = {\eta_{\mu \nu}} + {h_{\mu \nu}}
\end{equation}

with the condition that  $ \vert {h_{\mu \nu}} \vert \ll 1$.

Canonically using the above as perturbation and using the lowest order ( 
first order in $ {h_{\mu \nu}}$ ) approximation one does recover the NM
formulations. This is basically as per perturbation theory being valid for 
the weak field.

However as per our complementarity idea the metric for the case
under consideration, should break up into its geometric and topological 
components 

\begin{equation}
{g_{\mu \nu}} = {<{\eta_{\mu \nu}}>}_{geometry} + 
{<{h_{\mu \nu}}>}_{topology}
\end{equation}

which is basically  the ${\eta_{\mu \nu}}$ and the
${h_{\mu \nu}}$  terms respectively.
This is an exact result in our formulation.
However the perturbation idea would work ( within limits ) in as much as
$ \vert {h_{\mu \nu}} \vert \ll 1$
is actually small physically.
The exactness of our result makes sense as it turns out that there
are consistency issues when using perturbation theory in the above case
as emphasized by Wald [1 p 78]. To quote him
" One final, somewhat troublesome point deserves further comment. Above we 
showed that general relativity reduces to Newtonian gravity in an 
appropriate limit, but strictly speaking, we went beyond the linear 
approximation to show this. .. it illustrates the difficulties which 
occur when one tries to derive equations of motion from Einstein's 
equation via a perturbation expansion in the departure from flatness.
In order to obtain a good approximation to a solution 
to given order, one must use aspects of the higher order equations."
This just consolidates our assertion here that this is no perturbation 
actually.

\newpage

Given the success of the concepts introduced here we have formalized the 
same in terms of the the following theorem:

{\bf The Geometry-Topology Complementarity Theorem}:

{\bf Part A} Given any mathematical space, geometry and topology are 
exclusively and exhaustively, complementary aspects of its nature.

{\bf Part B} Under appropriate conditions, all physically measurable 
quantities should have geometric and topological parts as additive
quantities.

{\bf Part C} Parts A and B are generic in nature; ie. independent of the 
exact details of particular geometry and topology used in 
different models

The proof of Part A is already contained in the beginning para of this 
paper. The exclusiveness and exhaustiveness of the complementarity ensures 
Part B. The generality of the results implicit in Part A and B suggest
the correctness of Part C. In fact this theorem may be used to put 
constraints and demand consistency of various geometric and 
topological models used to describe a particular 
mathematical/physical reality.

Let us apply this Theorem to Einstein energy equation of 
Special Relativity

\begin{equation}
E^2 = {p^2}{c^2} + {{m_0}^2}{c^4}
\end{equation}

As the first part on right is clearly metric dependent - it 
constitutes the
geometric part of the equation. Hence clearly as per the above 
Theorem the rest mass $m_0$ is necessarily of topological origin.
As the theorem is of generic nature, it provides "existence" proofs.
It tells us about the nature of the rest mass without giving details of 
how and in what manner it arises. However as we are now aware of the
intrinsic topological nature of the rest mass, we should be able to
explore the same issue with greater clarity and confidence.

\newpage

\vskip 2 cm
\begin{center}
{\bf REFERENCES }
\end{center}

\vskip 2 cm

1. R M Wald, "General Relativity", Univ Chicago Press, Chicago (1984)

\vskip 2 cm

2. A Harvey and E Schucking, "Einstein's mistake and the 
cosmological constant", 
{\it Am J Phys}, {\bf 68} (2000) 723-727

\vskip 2 cm

3. E J Copeland, M Sami and S Tsujikawa, "Dynamics of Dark Energy"
{\it Int J Mod Phys}, {\bf D 15} (2006) 1753-1936

\vskip 2 cm

4. E I Guendelmann and A B Kaganowich, "Dark Energy and the Fifth Force 
Problem", {\it J Phys}, {\bf A 41} (2008) 164053

\vskip 2 cm

5. H Reichenbach, "The philosophy of space and time", Dover, 
New York (1957) (Original German edition in 1928)

\end{document}